\documentclass[pre,groupedaddress,showpacs,preprint]{revtex4}
\usepackage[english]{babel}
\usepackage{epsfig}
\usepackage{amsmath}
\usepackage{amssymb}
\usepackage{amsfonts}
\usepackage{array}
\usepackage{subfigure}
\usepackage{psfrag}

\DeclareMathOperator{\e}{e}

\begin{document}

\title{Thermally activated breakdown in a simple polymer model}

\author{S. Fugmann and I. M. Sokolov}

\affiliation{Institut f\"ur Physik,
Humboldt-Universit\"at zu Berlin,
Newtonstra\ss e 15, D-12489 Berlin, Germany}

\date{\today}

\begin{abstract}
We consider the thermally activated fragmentation of a homopolymer chain. In our simple model the dynamics of the intact chain is a Rouse one until a bond breaks and bond breakdown is considered as a first passage problem over a barrier to an absorbing boundary. Using the framework of the Wilemski-Fixman approximation we calculate activation times of individual bonds for free and grafted chains. We show that these times crucially depend on the length of the chain and the location of the bond yielding a minimum at the free chain ends. Theoretical findings are qualitatively confirmed by Brownian dynamics simulations.
\end{abstract}


\maketitle

The problem of thermally activated chain fragmentation (thermolysis) is of fundamental interest in understanding the degradation and stabilization properties of polymers \cite{Allen66}. Whenever the corresponding fragmentation kernel (i.e. the probability per unit time for a bond at a given position to break) is known as the function of the breakdown position, time, chain's length etc., the overall fragmentation process is well-described, and the distribution of fragments at whatever time can be obtained by the solution of the corresponding kinetic equation \cite{Ziff86_Macro}. Here the fragmentation kernel is the input into the universal theory, and many model forms for such kernels were formulated on the basis of parameterizing experimental observations or as simple analytical examples. However, up to our notion, the question of how does the corresponding kernel follows from the single polymer chain's dynamics was hardly considered. The only example known to us is contained in \cite{Hathorn01_MTS} which however deals with a model whose relation to the standard polymer dynamics ones is not quite evident. 

In what follows we discuss the thermally activated breakdown within a model which assumes that the dynamics of the chain is a Rouse one  \cite{Rouse53,DoiEdwards86} (as long as it doesn't break), i.e. we disregard excluded volume effects, which is a reasonable assumption in the case of tagged chains in melts and concentrated solutions, as well as hydrodynamical interactions. We moreover assume that that the breakdown of the bond (represented as the breakable harmonic spring) takes place as soon as its elongation achieves the preassigned value (equal for all bonds in the chain). The possibility of reestablishing the bond after breakdown (defect healing) is neglected. We consider the relevant situations of free and grafted chains. 

The thermally activated bond breakdown is essentially an example of an intrachain chemical reaction. However, contrary to e.g. polymer cyclization \cite{Doi75,Likthman06,Das08_PRL,Kolb97,Moreira04_JCP,Pastor96_JCP,Sokolov03_PRL} which was considered in quite a detail, this one was hardly tackled. The assumption that the bond breaks when achieving the given elongation simplifies the description, and corresponds to assuming the corresponding reaction to be purely diffusion-controlled reaction on a contact. Such problems can be casted into mathematical form of the first passage problem over a barrier to an absorbing boundary. Although the formulation of the overall problem is extremely simple, its solution is not, since the projection of the overall chain motion onto the reaction coordinate makes the corresponding diffusion strongly non-Markovian \cite{Szabo80_JCP}. This strongly non-Markovian nature of the problem involving multiple characteristic timescales reflects the fact that the reaction essentially takes place in a many-particle system.

In recent years theoretical methods have emerged to treat diffusion-controlled reactions among sites attached to polymers. Pioneering works go back to Wilemski and Fixman \cite{Wilemski74_1,Wilemski74_2} and conceptual advances were made by Doi \cite{Doi75}, de Gennes \cite{deGennes82} and others \cite{Szabo80_JCP}. But for all that, except for some special cases \cite{Likthman06}, the analytical theories of reaction diffusion in polymer physics fail to give an exact description of the reaction rates and rely on additional assumptions \cite{Pastor96_JCP}. However, as we proceed to show, the outcome of theoretical considerations within the framework of Wilemski and Fixman agrees qualitatively very good with the results obtained in Brownian dynamics simulations. Distinct from numerous studies on end-chain reactions \cite{Doi75,Likthman06,Das08_PRL,Kolb97,Moreira04_JCP,Pastor96_JCP,Sokolov03_PRL} or the studies on interior loop forming reactions \cite{Sung03_JCP}, we focus on the related but somewhat different problem of thermal activation of bonds, i.e., the first passage problem of nearest neighbor monomer distances. 

As we proceed to show the dissociation dynamics of the chain strongly depends on the location of the bond within the chain and the size of the system. Although the equilibrium distributions and activation barriers are the same for all the bonds, their activation times are not. At the free ends of the chain the first passage times are substantially lower compared to bonds in the middle of the chain (for a free chain) or at its grafted end (for a grafted one). Thus, the thermally activated fragmentation is expected to happen mostly at the chain ends. A similar behavior was found experimentally \cite{Madras96}. Furthermore it was shown that the forced rupture of adhesive contacts is strongly influenced by chain dynamics for undercritical forcing \cite{Barsegov08}.

The paper is structured as follows: In the next section we introduce our model and discuss its breakdown properties within the simple one-dimensional setup. We recall the dynamics of the underlying polymer model and in Sec.~\ref{s_dcr} an approach to calculate first passage times in diffusion controlled reactions based on Wilemski-Fixman approximation. In Sec.~\ref{s_rr} we study the impact of the location of the bond on its activation time for various chain lengths and activation energies. In Sec.~\ref{s_3dc} we generalize our approach to the experimentally relevant three-dimensional chain. Finally we summarize our results. 

\section{The model}
\label{s:model}

We first consider a one dimensional chain of $N+1$ monomers with coordinates $q_0$, ..., $q_N$. The interaction potential is given by
\begin{equation}
U(q_0,...,q_N)=\frac{1}{2}\kappa\sum_{i,j}\hat{R}_{i,j}q_iq_j\,,
\end{equation}
with the Rouse matrix $\hat{R}$. The overdamped dynamics of the beads follow the Langevin equations
\begin{equation}
\gamma \dot{q}_i=-\frac{\partial U}{\partial q_i}+\sqrt{2\gamma k_BT}\xi_i\,,
\label{eq:setLangevin1}
\end{equation}
with $\xi_i$ being independent delta-correlated Gaussian white noise, damping coefficient $\gamma$ and thermal energy $k_BT$. We consider the cases of a free chain as well as of a grafted chain with $q_0(t)\equiv 0$.  The Rouse matrix reads
\begin{equation}
\hat{R}=
\begin{pmatrix}
(2-\epsilon) & -1 & & & 0 \\
-1 & 2 & \ddots & & \\
 & \ddots & \ddots & &\\
 &  & & 2 & -1 \\
0 & & & -1 & 1
\end{pmatrix}\,,
\label{eq:rouseMatrix}
\end{equation}
with $\epsilon=1$ for the free chain and $\epsilon=0$ for the grafted one. We pass to a dimensionless time, $\tilde{t}=\kappa/\gamma t$, and neglect in the following the tilde in our notation. As long as the chain is intact, it is described by the standard Rouse model of polymer dynamics whose corresponding interaction strength reads $\kappa=3k_BT/b^2$ ($b^2$ being the mean squared length of a single bond) \cite{Rouse53,DoiEdwards86}.

\begin{figure}
\begin{center}
\includegraphics[height=6.5cm,width=9.cm]{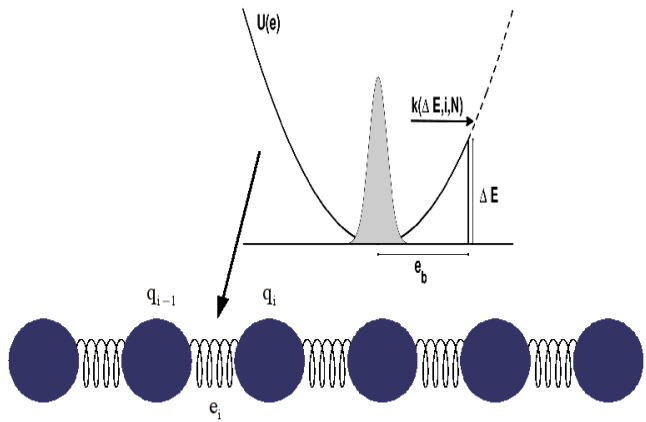}
\end{center}
\caption{\label{fig:sketch1d}One-dimensional chain of $N+1$ monomers connected by harmonic springs. Superimposed is the interaction potential with a barrier of height $\Delta E$ at $e_b=q_{i-1}-q_{i}$. The activation rate is assumed to depend on the barrier height, the position of the bond in the chain and the system size.}
\end{figure}

The system of coupled monomers is sketched in Fig.~\ref{fig:sketch1d} for a free chain. Superimposed is the equilibrium distribution of $e_i$,
\begin{equation}
\psi_{eq}(e_i)=\frac{1}{\sqrt{2\pi\phi_0}}\exp\left[-\frac{e_i^2}{2\phi_0}\right]\,,\mbox{with}\,\,\,\phi_0=\langle e_i^2\rangle\,,
\end{equation}
which is the same for all $i$.


The problem of thermally activated breakdown can be casted into a first passage problem of a reaction coordinate $e_i=q_i-q_{i-1}$ over a barrier of height $\Delta E=U(e_b)-U(0)$ situated at $e_b$. In our harmonic model the reaction is assumed to be irreversible and to take place once the reaction coordinate reaches $e_b$ in the harmonic potential well. Thus the absorbing boundary at $e_b$ introduces a cut-off of the harmonic potential. 

In order to study the thermolysis of the chain, all $N$ bonds are assumed to have a cut-off at $e_b$ and the chain is broken as soon as the first $e_i$ reaches the barrier. For the systematic study of the mean first passage times (the inverse activation rates) $\tau_{mfp}(i)$ of individual bonds  only one of them is breakable (the one under study, with a cut-off at $e_b$), the remaining $N-1$ bonds are described by perfect harmonic springs.

First, we consider numerically the thermolysis of the whole chain. Thus, the set of coupled equations (\ref{eq:setLangevin1}) was integrated by use of a Heun integration scheme. Averages were performed over an ensemble of at least $10000$ trajectories. Initial configurations were generated using the equilibrium distribution of the $e_i$.

\begin{figure}
\begin{center}
\subfigure[]
{\includegraphics[height=6.5cm,width=7.5cm]{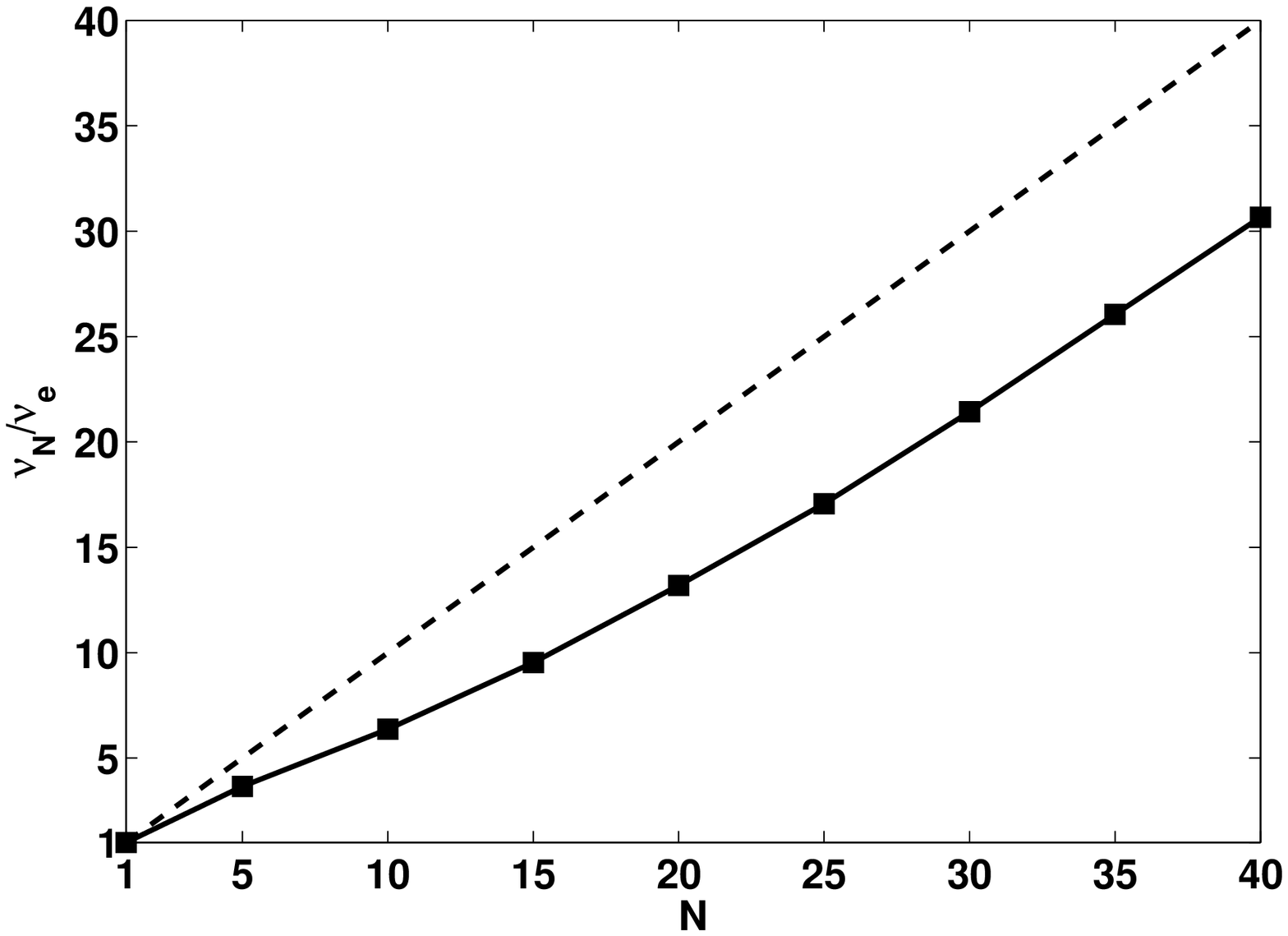}}
\subfigure[]
{\includegraphics[height=6.5cm,width=7.5cm]{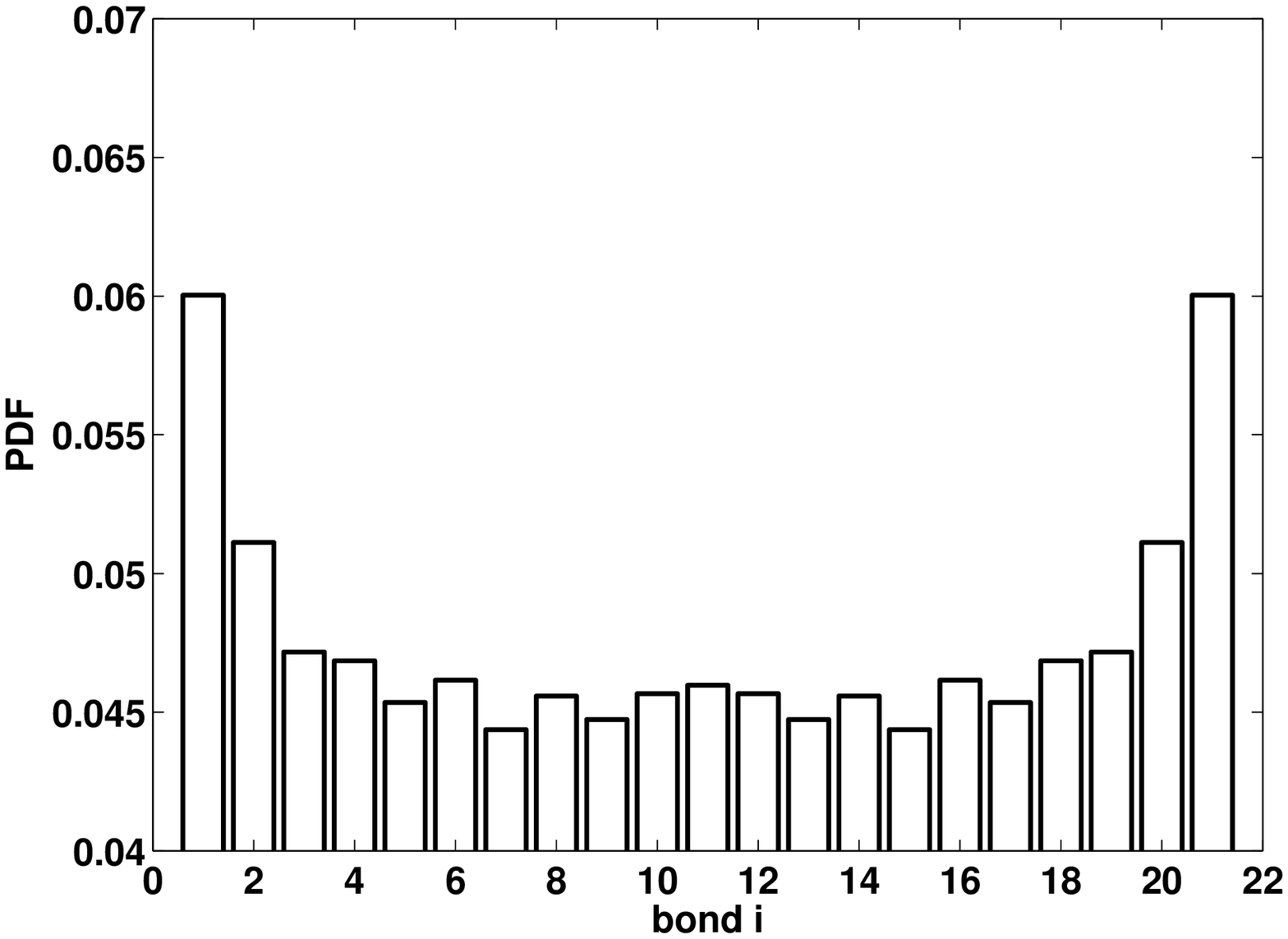}}
\end{center}
\caption{\label{fig:scalingRate}Panel (a): The fragmentation rate of the chain as a function of its length. Shown is $\nu_N/\nu_e$. Panel (b): Probability distribution of activation for a chain with $N=22$ bonds. The barrier height is $\Delta E/k_BT=5$.}
\end{figure}

The survival probability of a single bond in the chain is given by
\begin{equation}
W_i(t)=\exp\left[-\nu(i) t\right]\,,
\end{equation}
with the breakdown rate $\nu_i$ equal to the activation rate of the bond $i$ over the barrier, which on its turn is proportional to the inverse mean first passage time to $e_b$, i.e., $\nu(i)=1/\tau_{mfp}(i)$. The survival probability of the whole chain (assuming uncooperative activation of the bonds) is thus
\begin{equation}
W_N(t)=\prod_{i=1}^N W_i(t)=\exp\left[-\nu_N t\right]\,,
\end{equation}
with the fragmentation rate of the chain
\begin{equation}
\nu_N=\sum_{i=1}^N\nu(i)\,.
\end{equation}
For a set of $N$ bonds with equal activation rates $\nu(i)=\nu_e$ we have $\nu_N/\nu_e=N$. Due to the coupled dynamics this scaling with $N$ is shown not to hold true. In Fig.~\ref{fig:scalingRate} we depict the numerically obtained activation rates for free chains of different length. The scaling differs drastically from the linear one for small chains and approaches asymptotically a slope of almost one in the limit of long chains. The rate is always below its value for the case of identically activated bonds. We conclude, that in longer chains especially the inner bonds have lower activation rates, or---in turn---larger mean first passage times. Hence a chain is expected to be activated with higher probability close to its ends. This is shown in panel (b) of Fig.~\ref{fig:scalingRate} where we present the probability density distribution of activation as a function of the location of the bond in the chain.

The advantage of this simple model is that as long as the chain is intact, it is described by the standard Rouse model of polymer dynamics, and therefore represents sufficiently well what happens in melts and concentrated solutions.

\subsection{Normal modes of the free chain}

The set of equations \eqref{eq:setLangevin1} can be decoupled by transformation to normal coordinates \cite{DoiEdwards86}.
For the free chain the normalized eigenvectors (normal modes) of \eqref{eq:rouseMatrix} are
\begin{equation}
\hat{x}_k(i)=\sqrt{\frac{2}{N+1}}\cos\left(\left(i+\frac{1}{2}\right)\frac{k\pi}{N+1}\right)\,,
\end{equation}
and the corresponding eigenvalues can be found as
\begin{equation}
\lambda_k^f=4\sin\left(\frac{\pi}{2}\frac{k}{N+1}\right)^2\,.
\end{equation}
They are inverse proportional to the relaxation times of single modes. The decoupled equations of motions describe independent Ornstein-Uhlenbeck processes with relaxation times
\begin{equation}
\tau_k^f=\frac{1}{\lambda_k^f}
\label{eq:lamf}
\end{equation}
for $k=1,...,N$.

Representing the monomers motion with respect to the normal modes, the monomers coordinate is given by
\begin{equation}
q_i(t)=x_0+2\sum_{k=1}^N x_k(t)\cos\left(\left(i+\frac{1}{2}\right)\frac{k\pi}{N+1}\right)\,,
\label{eq:trafo1}
\end{equation}
with $\lbrace x_k(t)\rbrace$ being the new coordinates
\begin{equation}
x_k(t)=\frac{1}{N+1}\sum_{i=0}^N q_i(t)\cos\left(\left(i+\frac{1}{2}\right)\frac{k\pi}{N+1}\right)\,.
\label{eq:trafo2}
\end{equation}
Due to the choice of normalization the dynamics of $x_0$ represent the time-evolution of the chain's center of mass.

\subsection{Normal modes of the grafted chain}

For the grafted chain the normalized eigenvectors of \eqref{eq:rouseMatrix} are
\begin{equation}
\hat{x}_k(i)=\frac{2}{\sqrt{2N+1}}\sin\left(i\pi\frac{2k-1}{2N+1}\right)\,,
\end{equation}
and the corresponding eigenvalues can be found as
\begin{equation}
\lambda_k^g=4\sin\left(\frac{\pi}{2}\frac{2k-1}{2N+1}\right)^2\,.
\end{equation}
Representing the monomers motion with respect to the normal modes, the monomers coordinate is given by
\begin{equation}
q_i(t)=\frac{2}{\sqrt{2N+1}}\sum_{k=1}^N x_k(t)\sin\left(i\pi\frac{2k-1}{2N+1}\right)\,.
\label{eq:trafo3}
\end{equation}
The relaxation times of single modes are
\begin{equation}
\tau_k^g=\frac{1}{\lambda_k^g}
\label{eq:lamg}
\end{equation}
for $k=1,...,N$. Note that longest relaxation time of the grafted chain,$\tau_1^g$, is approximately four times larger than $\tau_1^f$.


\section{Thermolysis as diffusion-controlled reaction}
\label{s_dcr}

In order to calculate the mean first passage times we briefly recall an approach put forward by de Gennes \cite{deGennes82}, based on the pioneering work of Wilemski and Fixman \cite{Wilemski74_1,Wilemski74_2}. 
The probability distribution functions of the distances $e_i$ are defined as $\psi(e_i,t)$. Their dynamics follow the generalized reaction-diffusion equation:
\begin{equation}
\mathcal{L}\psi=-\mathcal{Q}\psi\,,
\label{eq:rd1}
\end{equation}
with $\mathcal{L}$ being the diffusion operator in the absence of reaction and $\mathcal{Q}$ is the sink operator describing the reaction. Here it is assumed that the presence of reaction does not affect the distribution of all other variable $e_j$, $j\neq i$.
Choosing the delta-function sink (the Smoluchowski sink), i.e., $\mathcal{Q}(e_i)=K\delta(e_i-e_b)$, we have in the limit of infinite sink strength, i.e., $K\rightarrow \infty$, an absorbing boundary at $e_b$. The aim is now to derive an expression for the mean first passage time over a barrier corresponding to the energy growth towards $e_b$ into this infinitely deep and steep ``adhesion well". Eq.~\eqref{eq:rd1} can be formally solved using the Green's function method. The propagator $\mathcal{G}$ follows from the solution of the equation
\begin{equation}
\mathcal{L}\mathcal{G}(e_i,e_i^0;t-t_0)=\delta(e_i-e_i^0)\delta(t-t_0)\,.
\label{eq:rd4}
\end{equation}
$\mathcal{G}$ is the conditional probability of finding the bond with elongation $e_i$ at time $t$ provided that it was at $e_i^0$ at $t_0$.
The formal solution of Eq.~\eqref{eq:rd1} reads
\begin{equation}
\psi(e_i,t)=\psi_{eq}(e_i)-\int_0^t dt_0\int de_i^0\mathcal{G}(e_i,e_i^0;t-t_0)\mathcal{Q}(e_i^0)\psi(e_i^0,t_0)\,.
\label{eq:rd5}
\end{equation}
The fraction of bonds that have not crossed the barrier at $e_b$ is given by $\rho(t)$ and obeys the following relation
\begin{equation}
\begin{split}
-\frac{d\rho(t)}{dt}&=\int de \mathcal{Q}(e_i)\psi(e_i,t)=K\psi(e_b,t)\\
&=K\psi_{eq}(e_b)-K^2\int_0^t dt_0\mathcal{G}(e_b,e_b;t-t_0)\psi(e_b,t_0)
\end{split}
\label{eq:rd6}
\end{equation}
$\mathcal{G}(t)=\mathcal{G}(e_b,e_b;t)$ is called the memory function. Taking the Laplace transform of Eq.~\eqref{eq:rd6} together with $\rho(0)=1$ and $K\rightarrow \infty$, we have
\begin{equation}
1-s\tilde{\rho}(s)=\frac{\psi_{eq}(e_b)}{s\tilde{\mathcal{G}}(s)}\,.
\label{eq:rd8}
\end{equation}
For times exceeding the longest relaxation time the memory function approaches the steady state distribution $\psi_e(e_b)$. Thus, one usually introduces a function which vanishes when $t\rightarrow\infty$
\begin{equation}
h(t)=\frac{\mathcal{G}(t)}{\psi_{eq}(e_b)}-1\,.
\label{eq:rd9}
\end{equation}
Eq.~\eqref{eq:rd8} then reads
\begin{equation}
\tilde{\rho}(s)=\frac{\tilde{h}(s)}{1+s\tilde{h}(s)}\,.
\label{eq:rd10}
\end{equation}
It was shown \cite{Doi75,Moreira04_JCP}, that the long-time rate constant $s^*$ can be obtained by finding the pole of the previous equation. Furthermore in the limit of large characteristic time scales (potential barrier height much larger than $k_BT$), $s^*$ is close to zero and the mean first passage time $\tau_{mfp}$ is approximately given by
\begin{equation}
\tau_{mfp}=\frac{1}{\vert s^*\vert}\simeq\tilde{h}(0)\,.
\label{eq:rd11}
\end{equation}
Since the propagator is related to the two point joint probability distribution of $e_i$ via  $\mathcal{G}(e_i,e_i^0;t)=\psi_i(e_i,e_i^0;t)/\psi_{eq}(e_i^0)$, the mean first passage time becomes
\begin{equation}
\tau_{mfp}(i)=\int_0^{\infty}\left(\frac{\psi_i(e_b,e_b;t)}{\psi_{eq}^2(e_b)}-1\right)dt\,.
\label{eq:rp2}
\end{equation}
Following \cite{Doi75} it is shown in App.~\ref{a:green} that for any one-dimensional harmonic chain the two point joint probability distribution is given by
\begin{equation}
\psi_i(e_i,e_i^0;t)=\frac{1}{2\pi\phi_0\sqrt{1-c_i(t)^2}}\exp\left[-\frac{e_i^2+(e_i^0)^2-2c_i(t)e_ie_i^0}{2\phi_0(1-c_i(t)^2)}\right]\,.
\label{eq:rp1}
\end{equation}
with the normalized autocorrelation function $c_i(t)=\langle e_i(t)e_i(0)\rangle/\langle e_i^2\rangle$ which we will specify in the next section.

\section{Results}
\label{s_rr}

\subsection{Free chain}

In the free chain the time-correlation function of $e_i=q_i-q_{i-1}$ is
\begin{equation}
\langle e_i(t)e_i(0)\rangle=\frac{2}{N+1}\frac{k_BT}{\kappa}\sum_{k=1}^N \e^{-t/\tau_k^f} \sin\left[\frac{k\pi i}{N+1}\right]^2\,,
\label{eq:e11}
\end{equation}
where we have used the fact that different modes are orthogonal. The normalized correlation function $c_i(t)$ is
\begin{equation}
c_i(t)=\frac{2}{N+1}\sum_{k=1}^N \e^{-t/\tau_k^f} \sin\left[\frac{k\pi i}{N+1}\right]^2\,,
\label{eq:e12}
\end{equation}
with $\tau_k^f$ taken from Eq.~\eqref{eq:lamf}. For $N\gg i$ the sum can be replaced by an integral, furthermore $N+1\simeq N$. Substituting $k$ by $l=k\pi/N$ we have
\begin{equation}
c_i(t)\simeq\frac{2}{\pi}\int_{0}^\pi dl \e^{-t l^2} \sin\left[l i\right]^2\,.
\label{eq:e13}
\end{equation}
Thus for $N\gg i$, i.e., for bonds close to the chain's end, the correlation function $c_i(t)$ is no function of the system size. The integral in Eq.~\eqref{eq:e13} can be solved in terms of error functions.

\begin{figure}
\begin{center}
\subfigure[]
{\includegraphics[height=6.5cm,width=7.5cm]{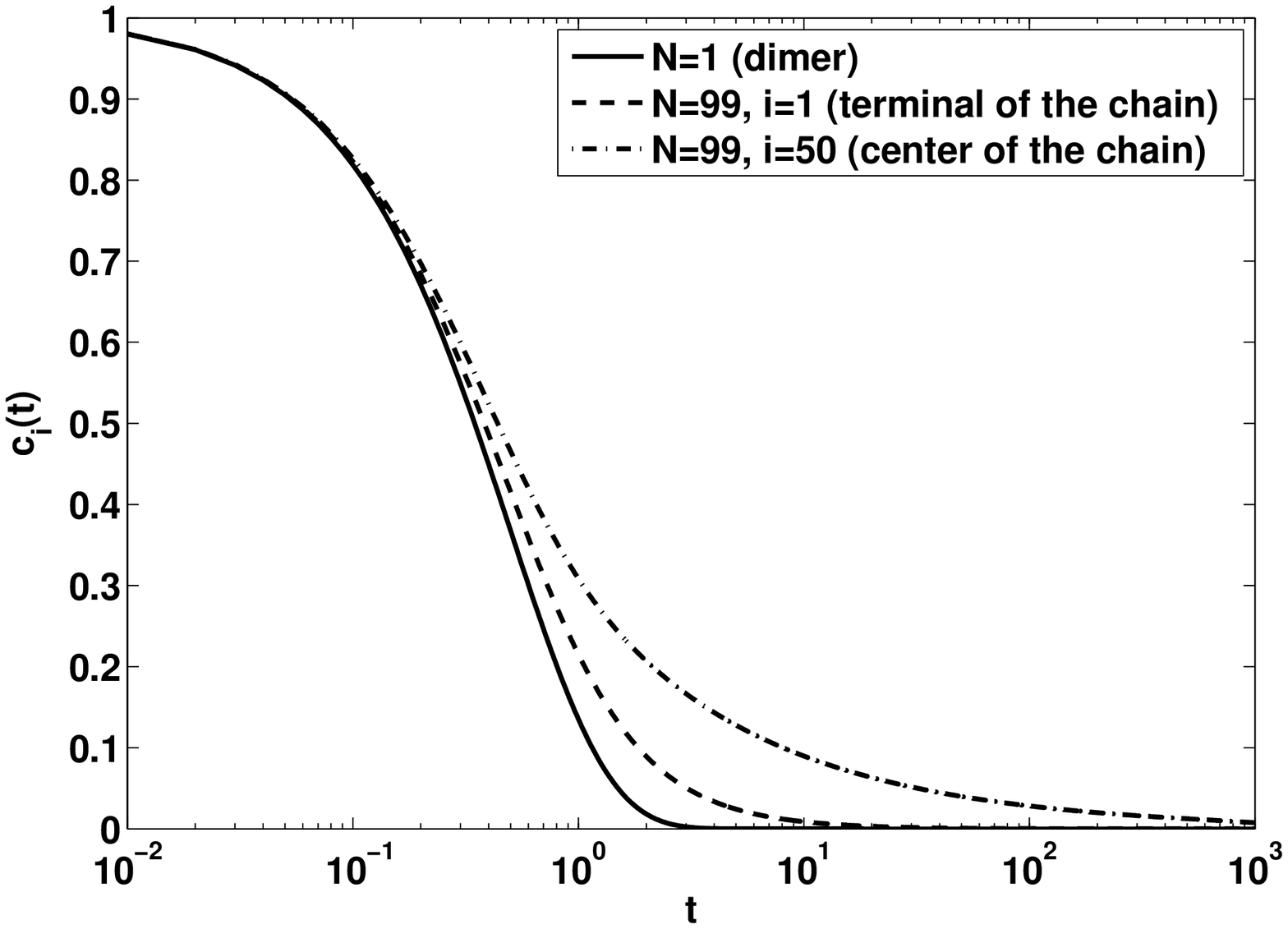}}
\subfigure[]
{\includegraphics[height=6.5cm,width=7.5cm]{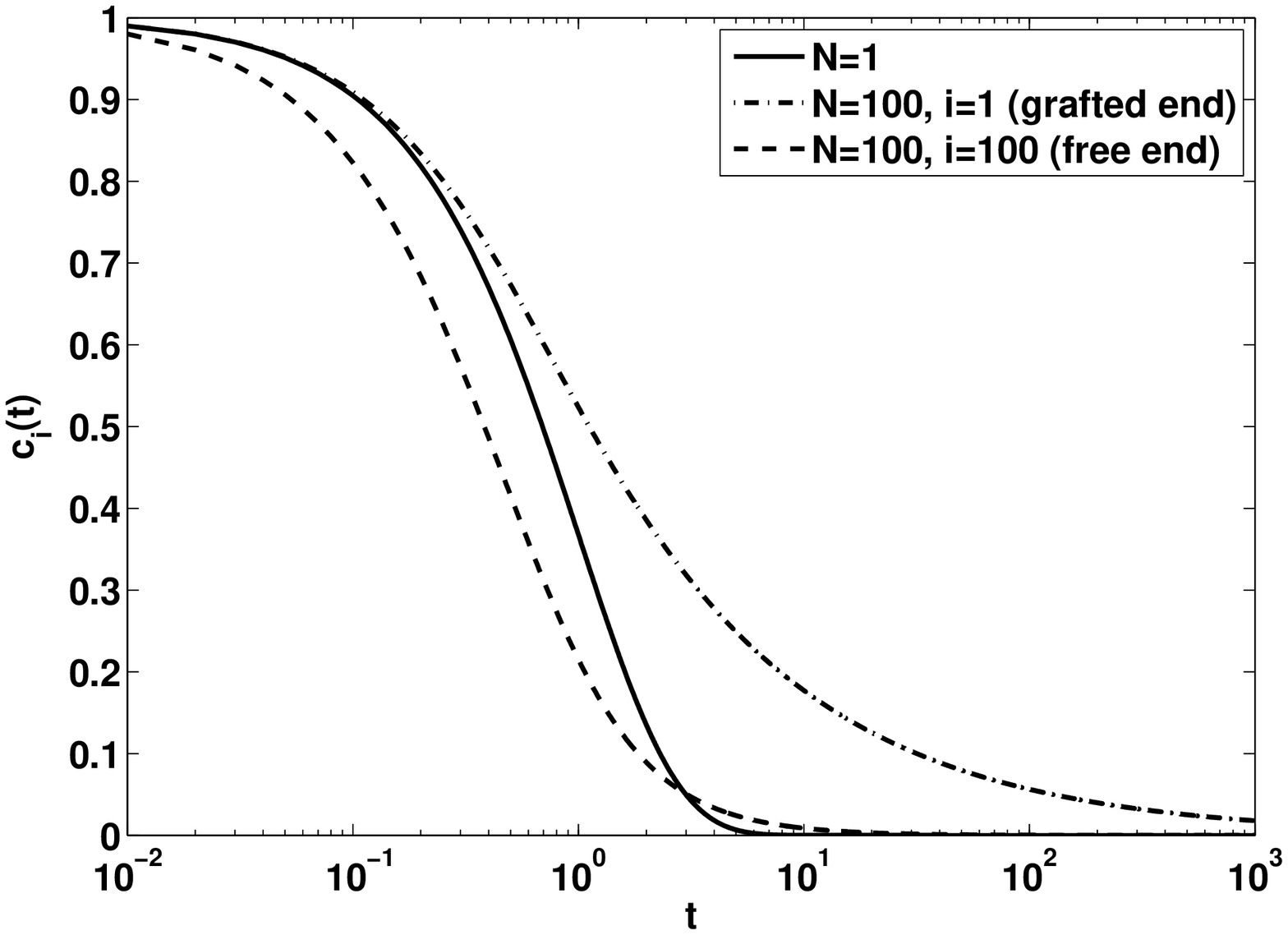}}
\caption{\label{fig:corr}Normalized autocorrelation function $c_i(t)$. Panel (a): Free chain with $99$ bonds. The correlation function is shown for $i=1$ (dashed line, terminal of the chain) and $i=50$ (dashed-dotted line, center of the chain). Superimposed is the correlation function of a dimer, i.e., a chain consisting of only one bond. Panel (b): Grafted chain with $100$ bonds. The correlation function is shown for $i=1$ (dashed-dotted line, grafted terminal of the chain) and $i=100$ (dashed line, free end of the chain). Superimposed is $c_1(t)$ for a single grafted monomer.}
\end{center}
\end{figure}

Since the mean first passage time is a functional of the correlation function, which by itself depends on the location of the bond in the chain, strong differences in the scaling of the correlation function might cause drastic differences of the activation rates (times) for bonds with different localization. Thus we first present in Fig.~\ref{fig:corr}, panel (a), the normalized autocorrelation function of a free chain of $99$ bonds. Depicted are the temporal correlation functions for a bond at one of the terminals of the chain ($i=1$, dashed line) and for a bond at the center of the chain ($i=50$, dashed-dotted line). Superimposed is the correlation function of a single bond, i.e., in a dimer (solid line). At short times ($t\ll\tau_1^f$) all curves coincide, while for longer times the correlations decrease much slower at the chain's center compared to the terminal where the correlation time (defined in the sense, that the correlation function is markedly different from zero) is also larger than for the dimer. Thus the correlation time is increased by orders of magnitude for bonds at the chain's center and the typical timescale of relaxation can reach and even overcome the mean activation timescale. Hence the dynamics become strongly non-Markovian.

\begin{figure}
\begin{center}
\subfigure[]
{\includegraphics[height=6.5cm,width=7.5cm]{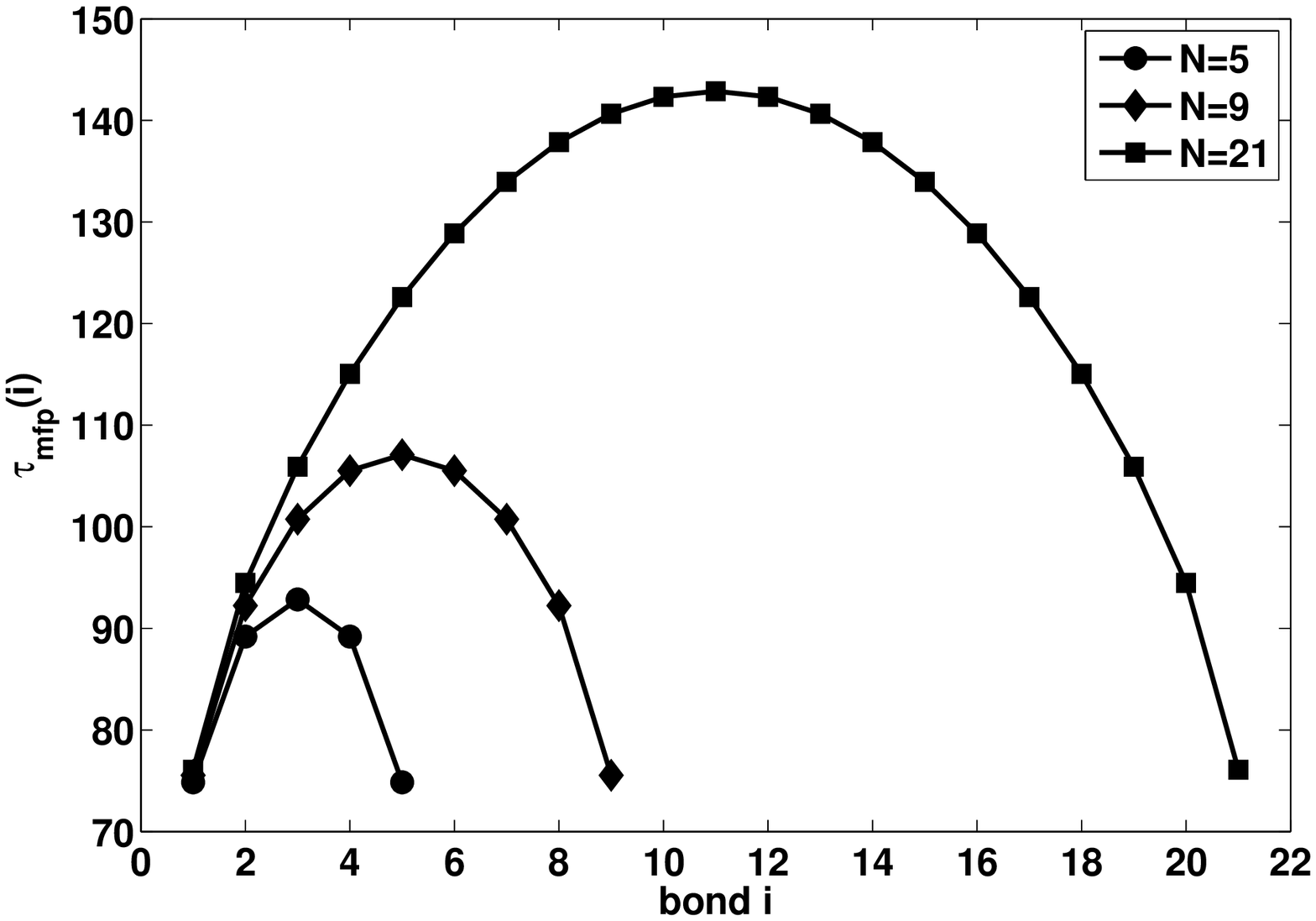}}
\subfigure[]
{\includegraphics[height=6.5cm,width=7.5cm]{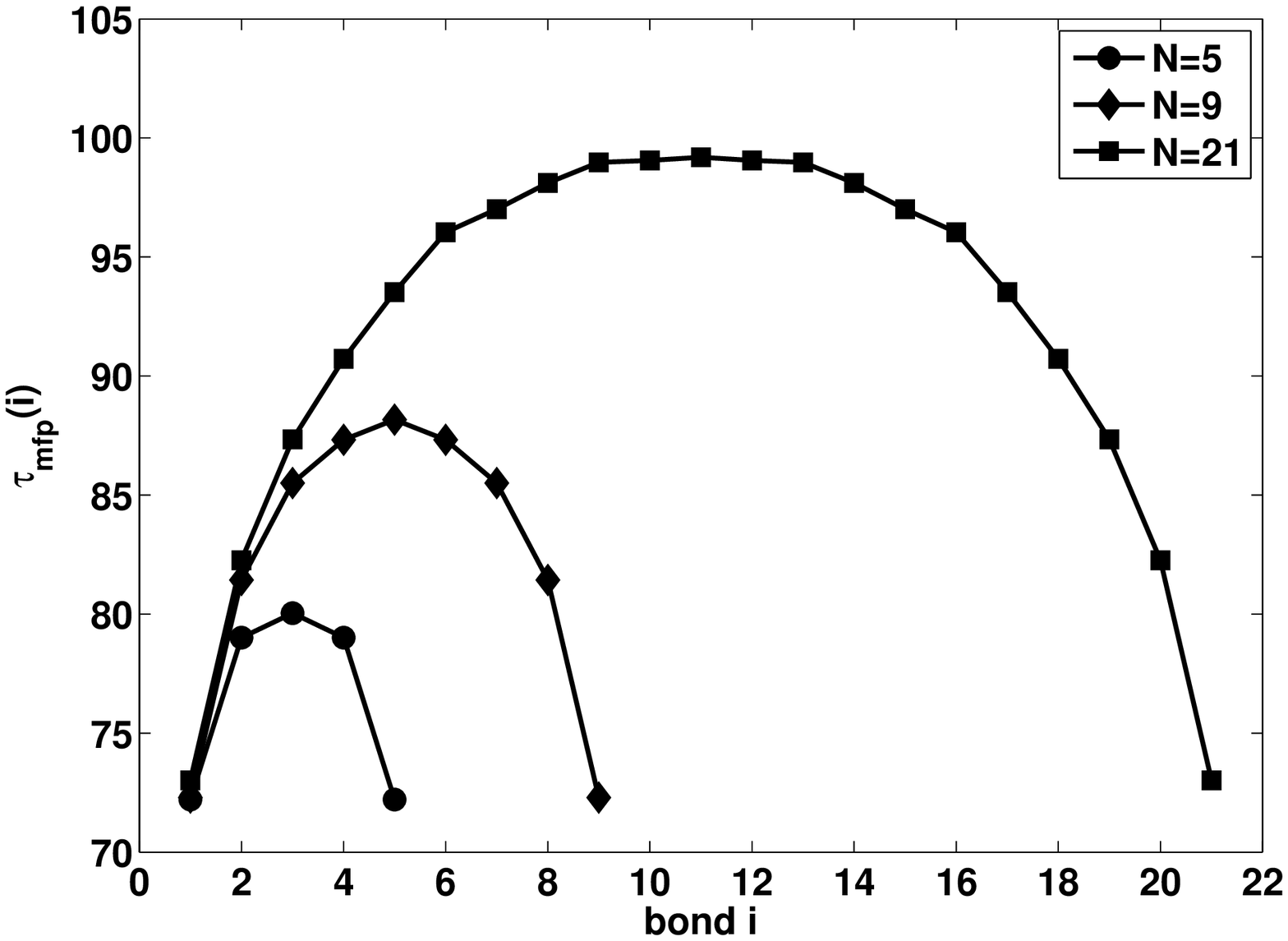}}
\caption{\label{fig:1dfree}Mean first passage time as a function of the bond position in the free chain. Panel (a): Mean first passage times obtained from Eq.~\eqref{eq:rp2} with the correlation function given in Eq.~\eqref{eq:e12}. Panel (b): Numerically obtained first passage times. The barrier height is $\Delta E=5k_BT$.}
\end{center}
\end{figure}

Since the analytical expression of the mean first passage time given by Eq.~\eqref{eq:rp2} is exact only in the Markovian limit, the theory is expected to work more accurate in predicting the times close to the chain's ends. However, as shown in Fig.~\ref{fig:1dfree}, it offers a qualitative picture that can explain observations in numerical simulations (i.e., it works qualitatively well in the whole range of $i$, which is typical for the Wilemski-Fixman approximation). In panel (a) we present the first passage times derived from Eq.~\eqref{eq:rp2} for three chain lengths and a barrier height of $\Delta E=5k_BT$. In panel (b) we show the outcome of Brownian dynamics simulations for the same set of parameter values. Qualitatively the outcome of the numerical simulations agrees very well with the theoretical prediction. For bonds at the ends of the chain the first passage times are smaller compared to the activation times of the inner bonds. For these bonds the theory is also quantitatively in good agreement with the numerical simulations. As expected the agreement becomes worse with enlarging distance from the terminals of the chain. However, both theory and simulations predict an increase of the activation time with increasing system size, even for the bonds located at $i=1$ and $i=N$. The observed effect is large already in relatively small systems (see panel (b) for $N=21$: $\tau_{mfp}(10)$ is about $40\%$ larger than $\tau_{mfp}(1)$) and becomes even larger in longer chains. Thus our study of the mean first passage times of individual bonds reveals that the activation times are smaller towards the chain ends what in turn cause there a higher probability of fragmentation. 

\begin{figure}
\includegraphics[height=7.cm,width=8.5cm]{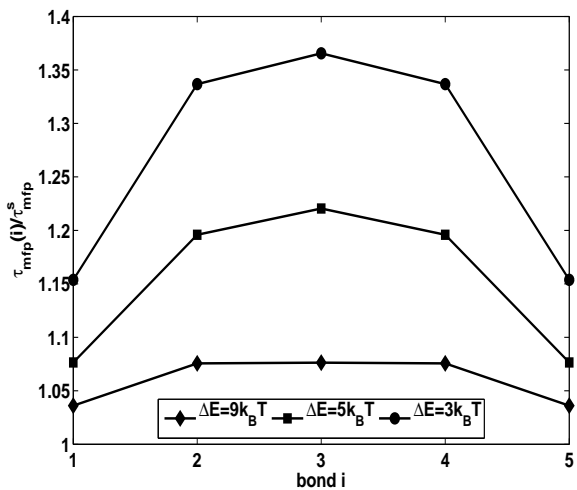}
\caption{\label{fig:1dcompf}Mean first passage times as a function of the position of the bond in the free chain for different values of the barrier height as given in the legend. The times are given in multiples of the first passage time in a single bond system (dimer). The chain has $N=5$ bonds.}
\end{figure}

The barrier height defines the intrinsic time scale of activation. In Fig.~\ref{fig:1dcompf} we depict the numerically obtained mean first passage times over barriers of different heights and a fixed chain length. The higher is the potential barrier the weaker is the increase of the times for the bonds at the center of the chain. This illustrates that the observed effect of the dependence of the dissociation time on the bond location is of highly non-Markovian nature. For lower barriers the activation times are comparable with the timescale of correlations in the chain and the dynamics is non-Markovian. We can conclude that the growth of activation times with enlarging distance from the chain's ends is the larger, the longer the chain and the lower the activation barrier is. 

\subsection{The grafted chain}

In the free chain we observed an increase of the activation times towards the central bond of the system. In the grafted chain there is also one free end, but another end is fixed. It was mentioned in the beginning that the longest relaxation time in the grafted chain (corresponding to the first normal mode) is four times longer compared to this time in the free chain. Thus we may expect that the non-Markovian aspect (induced by long correlation times) plays an even more important role in the barrier crossing dynamics.

In the grafted chain the normalized correlation of the $i$-th bond elongation is
\begin{equation}
c_i(t)=\frac{4}{(2N+1)}\sum_{k=1}^N \tau_k^g\e^{-t/\tau_k^g}\left(\sin\left(i\pi\frac{2k-1}{2N+1}\right)-\sin\left((i-1)\pi\frac{2k-1}{2N+1}\right)\right)^2\,,
\label{eq:ei1}
\end{equation}
with $\tau_k^g$ taken from Eq.~\eqref{eq:lamg}.

In Fig.~\ref{fig:corr}, panel (b), we show the normalized autocorrelation function as given in Eq.~\eqref{eq:ei1} of a grafted chain of $100$ bonds. We depict the temporal evolution of $c_i(t)$ for $i=1$, i.e., the bond at the grafted end (dashed-dotted line), and for $i=100$, i.e., the bond at the free terminal (dashed line). Superimposed is the correlation function for a single grafted bond. At short times the correlations drop down earlier compared to the single bond situation and the bond at the grafted end. Note that the dashed line of panel (b) coincides with the solid line in panel (a). Thus we conclude that for short times the dynamics of the bond at the free end resembles the dynamics of a dimer. We can infer, that for short times the dynamics of a bond at the chain's free terminal is the same, no matter whether the chain is fixed or not.

\begin{figure}
\begin{center}
\subfigure[]
{\includegraphics[height=6.5cm,width=7.5cm]{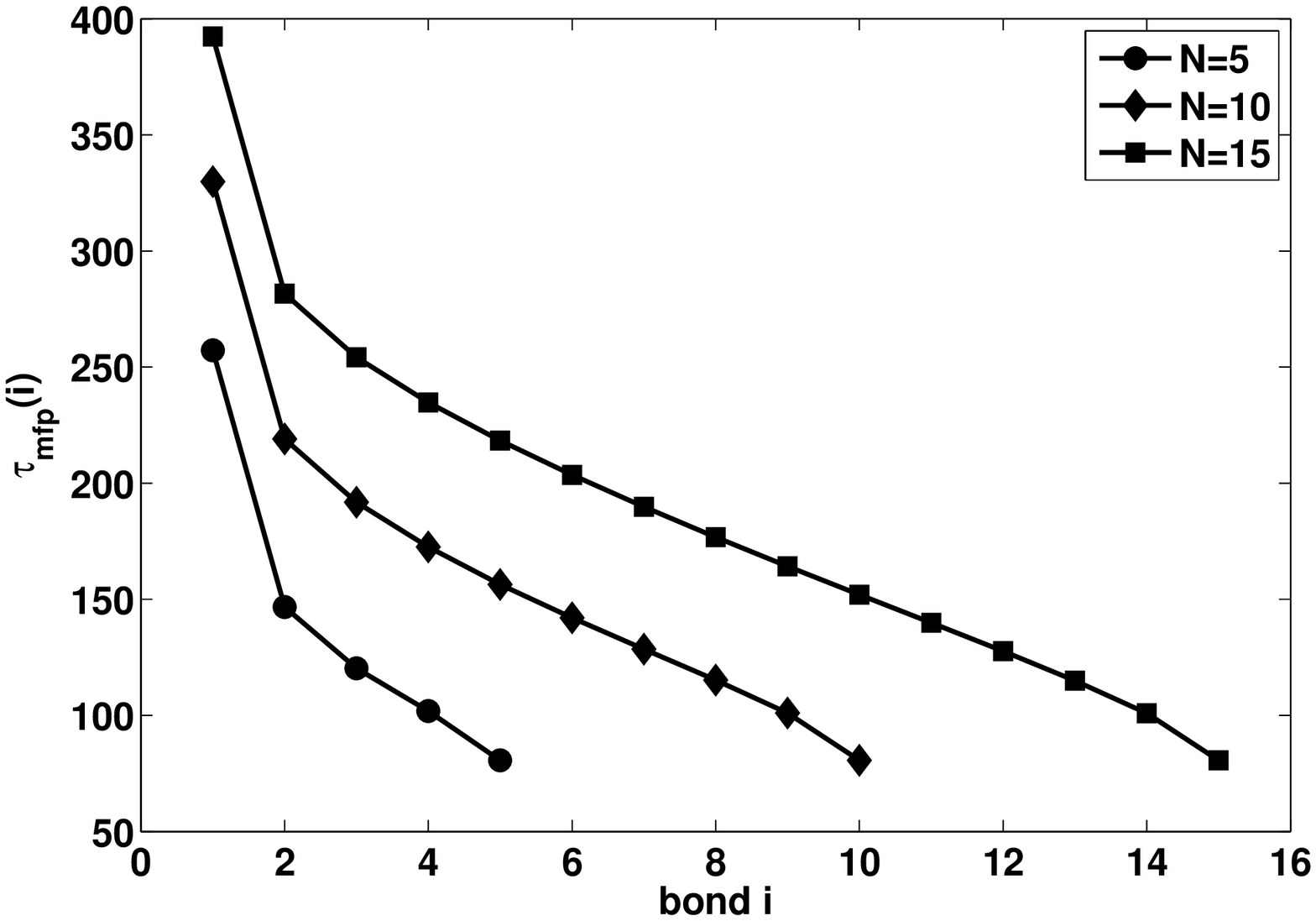}}
\subfigure[]
{\includegraphics[height=6.5cm,width=7.5cm]{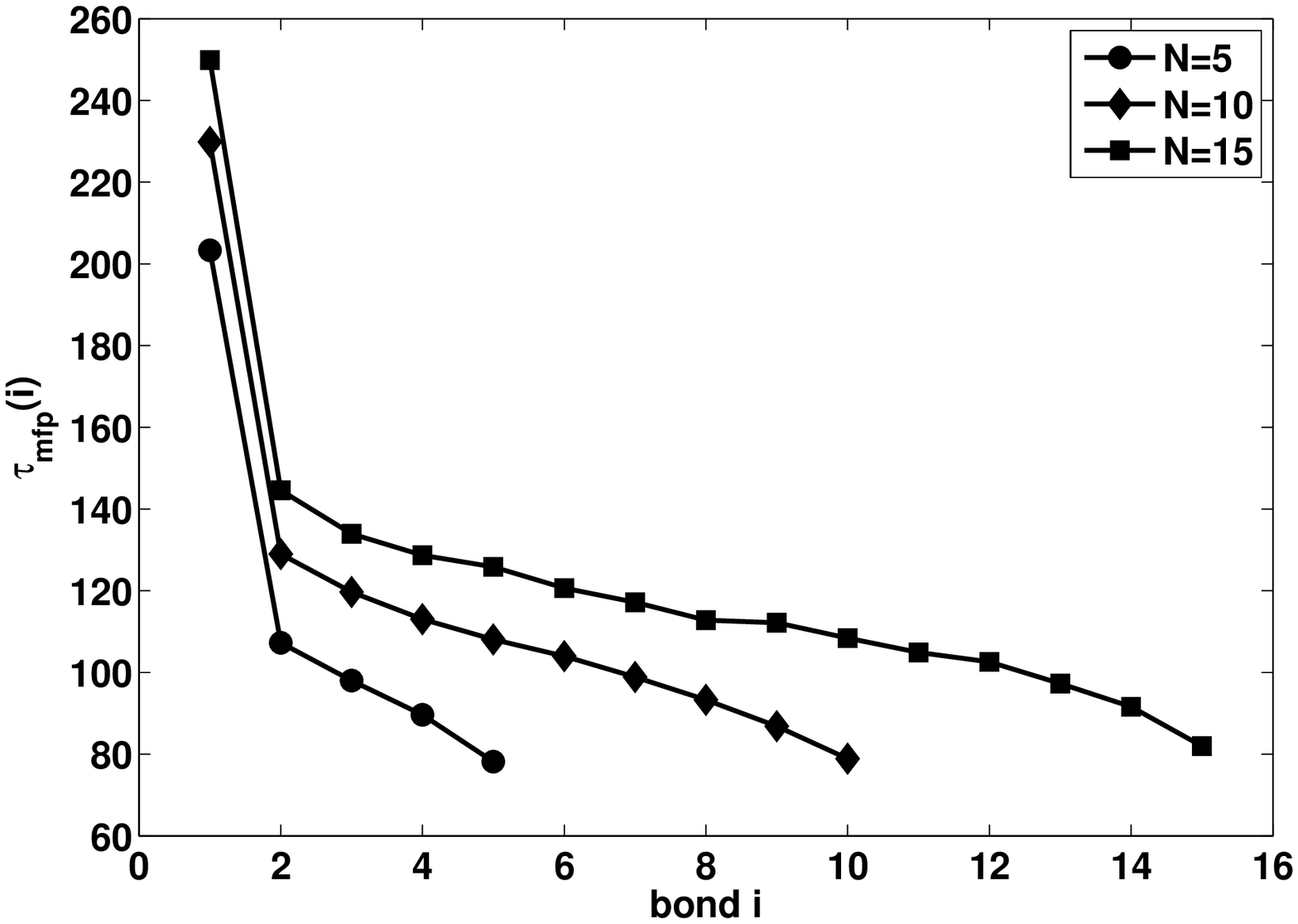}}
\caption{\label{fig:1dgrafted}Mean first passage time as a function of the bond position in the grafted chain. Panel (a): Mean first passage times obtained from Eq.~\eqref{eq:rp2} with the correlation function given in Eq.~\eqref{eq:ei1}. Panel (b): Numerically obtained first passage times. The barrier height is $\Delta E=5k_BT$.}
\end{center}
\end{figure}

In Fig.~\ref{fig:1dgrafted}, panel (a), we present the first passage times derived from Eq.~\eqref{eq:rp2} for three chain lengths and a barrier height of $\Delta E=5k_BT$. In panel (b) we show the outcome of Brownian dynamics simulations for the same set of parameter values. As in the free chain, for bonds at the loose terminal the first passage times are smaller compared to the activation times of the inner bonds. For these bonds the theory is also quantitatively in good agreement with the numerical simulations. The first passage times gradually increase with enlarged distance from the free end of the chain and grow substantially at the grafted terminal. The effect is overestimated by the theory. However, theory and numerical simulation are in good qualitative agreement.

\begin{figure}
\includegraphics[height=7.cm,width=8.5cm]{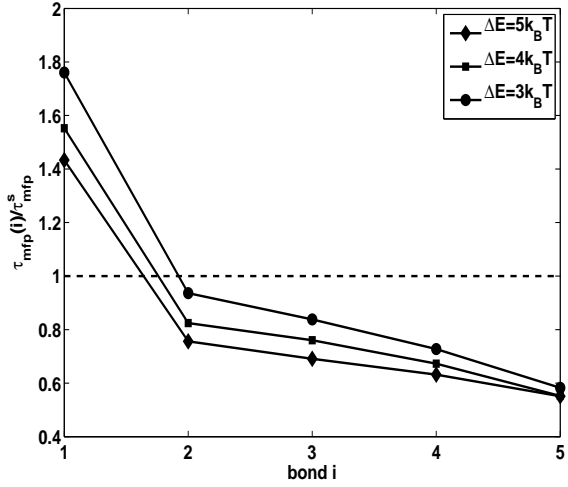}
\caption{\label{fig:1dcompg}Mean first passage times as a function of the position of the bond in the grafted chain for different values of the barrier height as given in the legend. The times are given in multiples of the first passage time in a single bond system. The chain consists of $N=5$ bonds.}
\end{figure}

As for the free chain we present in Fig.~\ref{fig:1dcompg} the numerically obtained mean first passage times over barriers of different height and a fixed chain length. The lower is the potential barrier, the stronger is the increase of the dissociation time along the chain. In the limit $\Delta E\gg k_BT$ $\tau_{mfp}(1)\rightarrow\tau_{mfp}^s$ while $\tau_{mfp}(i\neq 1)\rightarrow\tau_{mfp}^s/2$ (the value expected for a free dimer).

\section{The free 3D Rouse chain}
\label{s_3dc}

Let us complete our study and turn to the three-dimensional harmonic chain. $N+1$ beads are connected by $N$ harmonic springs. The ends are free and the chain's center of mass diffuses with an effective friction $\sim N$. In the three-dimensional chain system the two-point joint probability distribution of is given by
\begin{equation}
\psi_i(\textbf{e}_i,\textbf{e}_i^0;t)=\left(\frac{1}{2\pi\phi_0}\right)^{3}\frac{1}{\left(1-c_i(t)^2\right)^{3/2}}\exp\left[-\frac{1}{2\phi_0}\frac{\textbf{e}_i^2+(\textbf{e}_i^0)^2-2c_i(t)\textbf{e}_i\cdot\textbf{e}_i^0}{1-c_i(t)^2}\right]\,.
\label{eq:3d1}
\end{equation}
The derivation follows the same steps as shown in App.~\ref{a:green} and is given for example in \cite{Doi75, Sung03_JCP}.
After averaging over angles the distribution reads
\begin{equation}
\psi(e_i,e_i^0;t)=\frac{2e_ie_i^0}{\pi\phi_0^2c_i(t)\sqrt{1-c_i(t)^2}}\sinh\left[\frac{c_i(t)e_ie_i^0}{\phi_0\left(1-c_i(t)^2\right)}\right]\exp\left[-\frac{e_i^2+(e_i^0)^2}{2\phi_0\left(1-c_i(t)^2\right)}\right]\,.
\label{eq:3d2}
\end{equation}
Together with Eq.~\eqref{eq:rp2} the mean first passage time reads
\begin{equation}
\tau_{mfp}(i)=\int_0^{\infty}\left(\frac{\phi_0}{e_b^2c_i(t)\sqrt{1-c_i(t)^2}}\sinh\left[\frac{c_i(t)e_b^2}{\phi_0\left(1-c_i(t)^2\right)}\right]\exp\left[-\frac{e_b^2c_i(t)^2}{\phi_0\left(1-c_i(t)^2\right)}\right]-1\right)dt\,.
\label{eq:3d4}
\end{equation}
with $c_i(t)$ given in Eq.~\eqref{eq:e12}.
%

\begin{figure}
\includegraphics[height=7.cm,width=8.5cm]{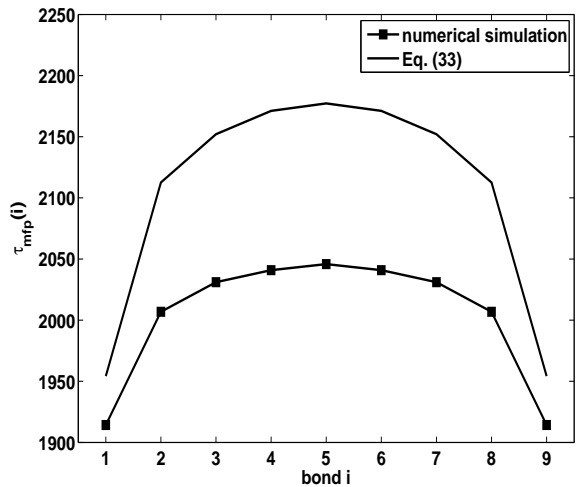}
\caption{\label{fig:3drates_de6_10}Mean first passage times as a function of the position of the bond in the chain. The barrier height is $\Delta E/k_B T=10$ . The chain length is $N=9$.}
\end{figure}

In Fig.~\ref{fig:3drates_de6_10} we compare the mean first passage times of $\vert \textbf{e}_i\vert$ derived from Eq.~\eqref{eq:3d4} and from Brownian dynamics simulations for a barrier height of $\Delta E/k_B T=10$. As in the one-dimensional system, there is a qualitative agreement between theory and numerical simulations. Even for a relatively short chain consisting of $N=9$ bonds, there is an increase of the activation time of about $8\%$, which will be even larger for longer chains and/or lower activation barriers which are of relevance in biological systems.

\section{Summary}
\label{s_sum}

Let us summarize our findings. We studied the thermally activated fragmentation of a homopolymer chain. It was shown that the fragmentation rate of the chain follows a nonlinear scaling as a function of the number of breakable bonds in the system. Bond breakage happens with higher probability at free chain ends. Studying the activation times of individual bonds we focused on the impact of their location in the chain and the length of the latter. It was found that towards the center of the free chain as well as towards the grafted terminal of the fixed chain the activation times increase substantially. The theoretically predicted effect is qualitatively confirmed by Brownian dynamics simulations both in one- and three-dimensional systems. The observed effects are large and therefore the framework presented in this article may help to interpret real experimental data.

\begin{acknowledgments}
The authors thankfully acknowledge financial support by DFG within the SFB 555 research collaboration program.
\end{acknowledgments}

\appendix
\section{Two-point joint probability distribution}
\label{a:green}

The two-point joint probability distribution is given by
\begin{equation}
\psi_i(e_i,e_i^0;t-t_0)=\idotsint d\textbf{q}d\textbf{q}_0\delta(e_i-(q_i-q_{i-1}))\delta(e_i^0-(q_i^0-q_{i-1}^0))\psi_0\left(\textbf{q},\textbf{q}_0;t-t_0\right)\,.
\label{eq:app1}
\end{equation}
Since $\psi_0$---the joint probability distribution of all of the chains coordinates---is a multivariate Gaussian, the integral in Eq.~\eqref{eq:app1} yields a multivariate Gaussian \cite{Wang45}. The general representation is
\begin{equation}
\psi_i(e_i,e_i^0;t)=N\exp\left[-\frac{1}{2}m_1(t)e_i^2-\frac{1}{2}m_2(t)(e_i^0)^2-\frac{1}{2}m_3(t)e_ie_i^0\right]\,,
\label{eq:app2}
\end{equation}
with $N$ being a normalization prefactor. From
\begin{equation}
\int de_i \psi_i(e_i,e_i^0;t)=\psi_{eq}(e_i^0)\,,
\label{eq:app3a}
\end{equation}
\begin{equation}
\int de_i^0 \psi_i(e_i,e_i^0;t)=\psi_{eq}(e_i)\,,
\label{eq:app3b}
\end{equation}
and the normalization restraint of the probability density distribution we obtain $m_1=m_2=m$ and $N=\sqrt{m^2-m_3^2/4}/(2\pi)$. We derive
\begin{equation}
\psi_{eq}(e_i)=\frac{\sqrt{m-\frac{m_3^2}{4m}}}{2\pi}\exp\left[-\left(m-\frac{m_3^2}{4m}\right)\frac{e_i^2}{2}\right]\,.
\label{eq:app4}
\end{equation}
Furthermore we derive $\langle e_i^2\rangle=\phi_0=1/(m-m_3^2/(4m))$.
From
\begin{equation}
c_i(t)=\frac{\phi_t}{\phi_0}=\frac{1}{\phi_0}\int de_ide_i^0\psi(e_i,e_i^0;t)e_ie_i^0\,,
\label{eq:app5}
\end{equation}
it follows that $c_i(t)=-m_3/(2m)$ and subsequently $1/m=\phi_0(1-c_i(t)^2)$. Finally we have
\begin{equation}
\psi_i(e_i,e_i^0;t)=\frac{1}{2\pi\phi_0\sqrt{1-c_i(t)^2}}\exp\left[-\frac{e_i^2+(e_i^0)^2-2c_i(t)e_ie_i^0}{2\phi_0(1-c_i(t)^2)}\right]\,.
\label{eq:app6}
\end{equation}

\bibliographystyle{apsrev}

\end{document}